\documentstyle[12pt]{article}

\newcommand{\be}{\begin{equation}}
\newcommand{\ee}{\end{equation}}
\newcommand{\dd}{\partial}
\newcommand{\bea}{\begin{eqnarray}}
\newcommand{\eea}{\end{eqnarray}}

\begin{document}
\baselineskip .25in
\newcommand{\numero}{hep-th/0001114}  

\newcommand{\titre}{On the classical connection between the WZWN model\\
and topological gauge theories with boundaries}
\newcommand{\auteura}{ } 
\newcommand{\auteurb}{Noureddine Mohammedi \footnote{e-mail:
nouri@celfi.phys.univ-tours.fr}}

\newcommand{\placea}{{{\it{Laboratoire de Math\'{e}matiques et Physique Th\'{e}orique
\footnote{CNRS UPRES-A 6083} \\
Universit\'{e} de Tours\\ Parc de Grandmont, F-37200 Tours, France.}}}}
\newcommand{\beq}{\begin{equation}}
\newcommand{\eeq}{\end{equation}}

\newcommand{\abstrait}
{It is shown, at the level of the classical action,  that 
the Wess-Zumino-Witten-Novikov model is equivalent to a combined BF theory
and a Chern-Simons action in the presence of a unique boundary term. This 
connection relies on the techniques of non-Abelian T-duality in non-linear sigma models.
We derive some consistency conditions whose various solutions lead to different
dual theories.  Particular attention is paid to the cases of the Lie algebras $SO\left(2,1\right)$
and $SO\left(2,1\right)\times SO\left(2,1\right)$. These are shown to yield three
dimensional gravity only if the BF term is ignored.}
\begin{titlepage}
\hfill \numero  \\
\vspace{.5in}
\begin{center}
{\large{\bf \titre }}
\bigskip \\ \auteura 
\bigskip  {$\,$}
\bigskip \\
\auteurb  
\bigskip \\ 
\placea  \bigskip \\
\vspace{.9 in} 
{\bf Abstract}
\end{center}
\abstrait
 \bigskip \\
\end{titlepage}
\newpage

\section{Introduction}

One of the remarkable features of pure three dimensional gravity
is its reformulation as a Chern-Simons theory of gauge connections \cite{anna,witty1}.
The precise relationship between the two theories is well understood
when the spacetime has no boundaries. However, since the discovery of 
the BTZ (Ba\~nados-Teitelboim-Zanelli) black hole solution of three
dimensional gravity \cite{btz}, there has been a new interest in this subject.
This is partly in an attempt to provide a statistical mechanical interpretation
of the black hole entropy \cite{carlip1,balachan,strom}. 
The main idea behind this approach is, roughly, to treat 
the horizon of the black hole as a two dimensional boundary of spacetime.
This argument is based on some quantum mechanical considerations of black holes.
As a consequence, one is forced to deal with a theory of gravity in the presence of 
boundaries.  Imposing then appropriate boundary conditions would lead to 
physical observables that live on the boundary. After quantization, these
extra degrees of freedom (which are absent if the manifold has no boundaries)
would correctly account for the black hole entropy.
\par
At this point one might wonder about the precise nature of the two 
dimensional theory that gives rise to these observables. It has been known for 
a long time that a Chern-Simons theory, defined on a compact surface,  is intimately
related to a two dimensional Wess-Zumino-Witten-Novikov (WZWN) model \cite{witty2}. This connection
is in fact made at the quantum level and links the physical Hilbert space of the 
Chern-Simons theory to the conformal blocks of the WZWN model. If the three
dimensional manifold is with boundaries, then the equivalence between the 
two theories relies on the breaking of the gauge symmetry of the Chern-Simons
theory at the boundary \cite{elitzur,seiberg}. However, in the treatement of three dimensional gravity
as a Chern-Simons theory with boundaries, one is forced to preserve gauge
invariance as this is indeed what replaces diffeomorphism invariance in gravity \cite{anna,witty1}.
\par
The usual way to proceed in maintaining gauge invariance is as follows: One imposes
some boundary conditions on the gauge fields. This requires then 
the introduction of a corresponding boundary action in order to have a well-defined variational 
problem. However, the combined action in the bulk and on the boundary is not gauge invariant.
A remedy for this problem consists in introducing group-valued fields $g$ living on the boundary.
The resulting theory, for very particular boundary conditions, is a Chern-Simons theory
{\it {coupled}} to a WZWN model \cite{carlip1,balachan,gupta}. 
Another point of view in arriving to this conclusion is 
presented in \cite{kunst}. The quantisation of the WZWN theory is then believed to account for 
the entropy of the black hole. It should be emphasised that different boundary conditions
yield different conformal field theories on the boundary \cite{alot}. 
A BRST treatment leads also to the same conclusions \cite{hwang}, where different gauge fixings
give different theories. 
\par
On the other hand, the conventional method in showing the equivalence of 
Chern-Simons gauge theory and the WZWN model follows a different path \cite{elitzur,seiberg}.
In this procedure one does not introduce (by hand) new dynamical degrees of freedom on the 
boundary, namely the WZWN fields.  In fact, one starts from the Chern-Simons action
\be
I_{\rm{SC}}\left(A\right)=-{k\over 4\pi}
\int_{\cal{M}}{\mathrm{d}}^3y\,\epsilon^{\mu\nu\rho}\,{\mathrm{tr}}
\left(A_\mu\dd_\nu A_\rho +{2\over 3}
A_\mu A_\nu A_\rho\right)
\label{CS}
\ee
We assume here that ${\cal{M}}={\bf {R\times\Sigma}}$, where ${\bf \Sigma}$ is a disc
whose boundary we denote $\dd{\bf \Sigma}$ and whose radial and angular coordinates 
are $r$ and $\theta$, respectively. The time coordinate $\tau$ parametrises ${\bf R}$
and $\epsilon^{\tau r\theta}=1$.
It is clear that gauge invariance, which holds only at the level of the partition function if ${\cal M}$ 
has no boundaries, is broken by the presence of the boundary ${\cal{\dd M}}={\bf R\times\dd{\bf\Sigma}}$. 
The only gauge invariance left is the one for which the gauge parameters reduce to the identity at the
boundary. 
Expanding the action we find
\be
I_{\rm{SC}}\left(A\right)=-{k\over 4\pi}
\int_{\cal{M}}{\mathrm{d}}^3y\,{\mathrm{tr}} \left(
2A_\tau F_{r\theta} -A_r\dd_\tau A_\theta + A_\theta\dd_\tau A_r\right)
+{k\over 4\pi}\int_{\cal{\dd M}}{\mathrm{d}}^2x\,{\mathrm{tr}}
\left(A_\theta A_\tau\right)\,\,\,,
\ee 
where $F_{r\theta}=\dd_r A_\theta -\dd_\theta A_r +\left[A_r\,,\,A_\theta\right]$.
\par
If one ignores the boundary action, then $A_\tau$ is a non-dynamical field 
which forces, in the bulk, the constraint $F_{r\theta}=0$. The solution to this zero curvature
condition is $A_{i}=L^{-1}\dd_i L$, $(i=r,\theta)$, for some group element $L$. Upon substitution,
the Chern-Simons action reduces to a WZWN model. It is not clear though how to justify the fact that 
the boundary term is not taken into account.  The usual given explanation resides in choosing a 
gauge for which $A_\tau=0$. However, if one treats $A_\tau$ in the same manner everywhere
then one has the constraint $A_\theta=0$ on the boundary. Solving simultaneously the bulk and 
the boundary constraints would put conditions on $L$ at the boundary.    
Furthermore, 
one might choose to impose, by hand, some other boundary conditions. 
This in turn introduces a boundary action involving the gauge fields $A_\mu$. Here also
different boundary conditions would yield different theories, not necessarily of the WZWN type.
In conclusion, the precise
connection between Chern-Simons theory, on a manifold with boundaries, and the WZWN model
is far from being transparent.
\par
We notice that in all the above mentioned methods, the starting point is the Chern-Simons
theory. The WZWN model is obtained  either by a direct coupling or by a particular parametrisation 
of the gauge fields.  
The aim of this paper is to clarify, at the level of the action and in a classical manner, 
the nature of the relationship between the WZWN model and Chern-Simons theory. 
Our starting point is the WZWN model itself.
We find that
the WZWN theory is dual to a topological BF theory coupled to a Chern-Simons theory in the
bulk. The boundary action is unique for this equivalence to hold and no 
boundary conditions are imposed by hand. This is the same guiding
principle for possible boundary terms as that presented in \cite{teitel}.   Our approach makes use
of the techniques of non-Abelian T-duality transformations in non-linear sigma models \cite{tduality}. 
Here one relies on the existence of isometries and their gauging. In the case of the WZWN 
model there are two types of isometries. The first consists of the isometries for which 
the gauge fields live entirely on the boundary and are at most quadratic in the action.
The second corresponds to those isometries for which the gauge fields live on the boundary as well 
as on the bulk. It is this last category of isometries which is used in this study.
It has the advantage of allowing for a first order formulation of the WZWN model in terms
of gauge fields and Lagrange multipliers where the original fields do not appear anymore.          
Finally, we specialise to the case of the Lie algebras $SO\left(2,1\right)$ and 
$SO\left(2,1\right)\times SO\left(2,1\right)$ and explore their relation to three
dimensional gravity. We show that, in general, the two theories do not coincide. 
This is due to the presence of the BF theory action which is usually ignored
in the literature when comparing three dimensional gravity to the WZWN model.

\section{First order formulation of the WZWN model}

The action for the WZWN model defined 
on the group manifold ${\mathcal{M}}_{\mathcal{G}}$, based on the 
Lie algebra $\cal{G}$, is given by
\bea
S_{\rm{WZWN}}\left(g\right)&=&{k\over 8\pi}\int_{\dd \cal{M}} {\mathrm{d}}^2x\,
\sqrt{|\gamma |}\gamma^{\mu\nu}\,{\mathrm{tr}}\left(g^{-1}\dd_\mu
g\right)\left(g^{-1}\dd_\nu g\right)+\Gamma\left(
g\right) \nonumber\\
\Gamma\left(g\right)&=&{k\over 12\pi}\int_{\cal{M}} {\mathrm{d}}^3y\,
\epsilon^{\mu\nu\rho} \,{\mathrm{tr}}
\left(g^{-1}\dd_\mu g\right)\left(g^{-1}\dd_\nu g\right)
\left(g^{-1}\dd_\rho g\right)\,\,\,\,
\label{wzwn}
\eea
where $g\in {\mathcal{M}}_{\mathcal{G}}$ and $\cal{M}$ is a three dimensional
ball whose boundary is the two dimensional surface $\dd \cal{M}$.
The metric on this two dimensional  worldsheet is denoted by $\gamma_{\mu\nu}$.
The remarkable thing about this action is that a variation 
of the type $g\longrightarrow g+\delta g$ leads to a change in the action, $\delta S_{\rm{WZWN}}$, 
which is an integral over the boundary $\dd{\cal M}$ only. The equations of motion are obtained
without a need to impose any boundary conditions on the field $g$ or the 
variation $\delta g$. This property will serve as a guiding principle for our analyses
when gauge fields are included. We shall
adopt the philosophy of not imposing any boundary conditions on the fields but rather
let the equations of motion determine the behaviour of the fields everywhere. 
This point will be explained in details below.   
\par
As it is well-known, the WZWN action has the global symmetry
\be
g \longrightarrow LgR \,\,\,,
\label{trans}
\ee
where $L$ and $R$ are two constant (more precisely chiral) group elements. 
Our first step in constructing the dual theory is to gauge this symmetry. 
We, therefore, 
introduce two Lie algebra-valued gauge functions 
$A_\mu$ and $\widetilde{A}_\mu$ transforming as
\bea
A_\mu &\longrightarrow & LA_\mu L^{-1} - \dd_\mu LL^{-1}
\nonumber\\
\widetilde{A}_\mu &\longrightarrow & R^{-1} 
\widetilde{A}_\mu R + R^{-1}\dd_\mu R\,\,\,.
\label{gaugetrans}
\eea
Since the usual minimal coupling of the gauge fields does not lead
to an invariant theory, the gauged WZWN action is found by applying 
Noether's method \cite{ian,chris}. The final result is
\bea
S_{{\rm gauge}}&=& S_{\rm{WZWN}}\left(g\right) 
+ I_{\rm{SC}}\left(A\right)
-I_{\rm{SC}}\left(\widetilde{A}\right)\nonumber\\  
&+&{k\over 4\pi}\int_{\dd \cal{M}}{\mathrm{d}}^2x\,
{\mathrm{tr}}\left[P^{\mu\nu}_+\left(\dd_\mu gg^{-1} A_\nu
\right) -P^{\mu\nu}_-\left(g^{-1}\dd_\mu g \widetilde{A}_\nu\right)
-P^{\mu\nu}_-\left(A_\mu g\widetilde{A}_\nu g^{-1}\right)\right]
\nonumber\\
&+& {k\over 8\pi}\int_{\dd \cal{M}}{\mathrm{d}}^2x\,
\sqrt{|\gamma |}\gamma^{\mu\nu}\,{\mathrm{tr}}
\left(A_\mu A_\nu + \widetilde{A}_\mu \widetilde{A}_\nu\right)\,\,\,\,\,.
\label{gaugedwzwn}
\eea
We have defined, for convenience, the two quantities 
$ P^{\mu\nu}_{\pm}= \sqrt{|\gamma |}\gamma^{\mu\nu} \pm 
\epsilon^{\mu\nu}$. Notice also the natural appearance of 
the Chern-Simons actions corresponding to the gauge fields
$A_\mu$ and $\widetilde{A}_\mu$.
\par
There are only two 
kinds of subgroups of the transformations 
(\ref{trans}) for which the gauge fields live entirely on the boundary
$\dd\cal{M}$. This corresponds to the situation when the combination 
$\left[I_{\rm{CS}}\left(A\right)-I_{\rm{SC}}
\left(\widetilde{A}\right)\right]$ vanishes. The first category are the
diagonal subgroups where $R=L^{-1}$
and $A_\mu=\widetilde{A}_\mu$ with $L$ and $R$ being Abelian or non-Abelian 
group elements. 
The second kind are the axial subgroups for which $R=L$ and $A_\mu=-\widetilde{A}_\mu$,
where both $R$ and $L$ are Abelian group elements.
The 
sort of gauging we are interested in corresponds to taking 
the gauge functions $L$ and $R$ to be two independent, Abelian or  non-Abelian, group elements. 
The Chern-Simons parts of the gauged WZWN action (\ref{gaugedwzwn}) are 
then present.
\par
The next step towards the dual theory consists in casting the WZWN action in
a first order formulation. We begin from the following gauge invariant action 
\be
S_{\rm{total}}=S_{{\rm gauge}}
-{k\over 4\pi}
\int_{\cal{M}}{\mathrm{d}}^3y\,\epsilon^{\mu\nu\rho}\,{\mathrm{tr}}\left(B_\mu 
F_{\nu\rho}\right)
-{k\over 4\pi}
\int_{\cal{M}}{\mathrm{d}}^3y\,\epsilon^{\mu\nu\rho}\,{\mathrm{tr}}\left(\widetilde{B}_\mu 
\widetilde{F}_{\nu\rho}\right)\,\,\,\,.
\label{total}
\ee
Here $F_{\mu\nu}$ 
and $\widetilde{F}_{\mu\nu}$ are the two gauge curvatures correponding, respectively, 
to $A_\mu$ and $\widetilde{A}_\mu$. The Lie algebra-valued fields $B_\mu$
and $\widetilde{B}_\mu$ are two Lagrange multipliers 
transforming as $B_\mu\longrightarrow LB_\mu L^{-1}$ and
$\widetilde{B}_\mu\longrightarrow R^{-1}\widetilde{B}_\mu R$.
The equations of motion of these Lagrange multipliers (or their integration out
in a path integral formulation) 
lead to the constraints $F_{\mu\nu}=\widetilde{F}_{\mu\nu}=0$. The solutions to these two
equations are, up to gauge transformations, given by $A_\mu=h^{-1}\dd_\mu h$ and 
$\widetilde{A}_\mu=\widetilde{h}^{-1}\dd_\mu \widetilde{h}$ for two group elements $h$ and
$\widetilde{h}$. Substituting for $A_\mu$ 
and $\widetilde{A}_\mu$ in (\ref{total}) we find that $S_{{\rm total}}=
S_{\rm{WZWN}}\left(hg\widetilde{h}^{-1}\right)$.  Therefore, 
by a change of variables such that $g'=hg\widetilde{h}^{-1}$
(or equivalently by fixing a gauge such that
$h=\widetilde{h}=1$) one recovers the original WZWN model. We conclude
that the WZWN model in (\ref{wzwn}) is equivalent to the theory
described by the action in (\ref{total}).
\par
The gauge invariance of the action $S_{{\rm total}}$
allows one to choose a gauge such that $g=1$. 
Notice that this 
gauge choice is not possible for diagonal or axial subgroups. 
Substituting for $g$, we obtain the sought first order action
\bea
S_{{\rm first}}&=&  
I_{\rm{SC}}\left(A\right)
-I_{\rm{SC}}\left(\widetilde{A}\right)
-{k\over 4\pi}
\int_{\cal{M}}{\mathrm{d}}^3y\,\epsilon^{\mu\nu\rho}\,{\mathrm{tr}}\left(B_\mu 
F_{\nu\rho}\right)
-{k\over 4\pi}
\int_{\cal{M}}{\mathrm{d}}^3y\,\epsilon^{\mu\nu\rho}\,{\mathrm{tr}}\left(\widetilde{B}_\mu 
\widetilde{F}_{\nu\rho}\right) \nonumber\\
&-&{k\over 4\pi}\int_{\dd \cal{M}}{\mathrm{d}}^2x\,
P^{\mu\nu}_-\,{\mathrm{tr}}\left(A_\mu \widetilde{A}_\nu \right)
+{k\over 8\pi}\int_{\dd \cal{M}}{\mathrm{d}}^2x\,
\sqrt{|\gamma |}\gamma^{\mu\nu}\,{\mathrm{tr}}
\left(A_\mu A_\nu + \widetilde{A}_\mu \widetilde{A}_\nu\right)\,\,\,\,\,.
\label{first}
\eea
We should mention that the integration over the Lagrange multipliers
would always lead to the WZWN model. This follows from the above explanation
upon setting $g=1$, where we obtain $S_{\rm first}=S_{\rm{WZWN}}
\left(g'\right)$ with $g'=h\widetilde{h}^{-1}$. 
At this point, it is important to emphasise the crucial r\^ole played by the 
Lagrange multiplier terms in relating the first order action 
(\ref{first}) to the WZWN theory.  
A further practical manipulation consists in 
making a field redefinition such that $Q_\mu=B_\mu+{1\over 2}A_\mu$ and 
$\widetilde{Q}_\mu=\widetilde{B}_\mu-{1\over 2}\widetilde{A}_\mu$. The first order action
takes then the form 
\bea
S_{{\rm first}}&=&  
-{k\over 4\pi}
\int_{\cal{M}}{\mathrm{d}}^3y\,\epsilon^{\mu\nu\rho}\,{\mathrm{tr}}\left[Q_\mu 
F_{\nu\rho} + \widetilde{Q}_\mu \widetilde{F}_{\nu\rho}
-{1\over 3}A_\mu A_\nu A_\rho
+{1\over 3}\widetilde{A}_\mu \widetilde{A}_\nu \widetilde{A}_\rho
\right] \nonumber\\
&-&{k\over 4\pi}\int_{\dd \cal{M}}{\mathrm{d}}^2x\,
P^{\mu\nu}_-\,{\mathrm{tr}}\left(A_\mu \widetilde{A}_\nu \right)
+{k\over 8\pi}\int_{\dd \cal{M}}{\mathrm{d}}^2x\,
\sqrt{|\gamma |}\gamma^{\mu\nu}\,{\mathrm{tr}}
\left(A_\mu A_\nu + \widetilde{A}_\mu \widetilde{A}_\nu\right)\,\,\,\,\,.
\label{modfirst}
\eea
In this reformulation the equivalence to the WZWN model is much more transparent
and the distinction from pure Chern-Simons theory is evident.    
\par
Our first order action (\ref{first}) has a rich structure in terms of 
symmetries. Indeed,
the Lagrange multiplier
terms present in this action correspond to what is commonly known as 
``$BF$ theories". They are topological theories which
have been widely studied (see \cite{bf} for a review). A typical characteristic of these theories
is their invariance under the finite transformations
\bea
&&B_\mu\longrightarrow B_\mu +{\cal D}_\mu\alpha
=
B_\mu +  \dd_\mu \alpha 
+\left[A_\mu\,\,,\,\,\alpha\right]\nonumber\\
&&\widetilde{B}_\mu\longrightarrow \widetilde{B}_\mu +  
\widetilde{{\cal D}}_\mu\widetilde{\alpha}=
\widetilde{B}_\mu +  
\dd_\mu \widetilde{\alpha} 
+\left[\widetilde{A}_\mu\,\,,\,\,\widetilde{\alpha}\right]
\label{extratrans}
\eea
for some two arbitrary local functions $\alpha$ and $\widetilde{\alpha}$
evaluated in the Lie algebra $\cal{G}$.
In our case, this invariance does not hold due to the presence 
of the boundary. There are, however, two possible ways to restore this invariance.
The simplest would be to demande that $\alpha$
and $\widetilde{\alpha}$ vanish when evaluated at the boundary $\dd\cal{M}$.
The second is to put no restrictions on $\alpha$ and $\widetilde{\alpha}$ 
and to supply the first order action (\ref{first}) with
the additional boundary term
\be
S_{\rm {add}}=
-{k\over 4\pi}
\int_{\dd{\cal{M}}}{\mathrm{d}}^2x\,\epsilon^{\mu\nu}\,{\mathrm{tr}}\left(\lambda 
F_{\mu\nu}\right)
-{k\over 4\pi}
\int_{\dd{\cal{M}}}{\mathrm{d}}^2x\,\epsilon^{\mu\nu}\,{\mathrm{tr}}\left(
\widetilde{\lambda} 
\widetilde{F}_{\mu\nu}\right)\,\,\,.
\label{addaction} 
\ee
We associate then to the new fields $\lambda$ and $\widetilde{\lambda}$
the transformations $\lambda\longrightarrow \lambda -\alpha$ and
$\widetilde{\lambda}\longrightarrow \widetilde{\lambda} -\widetilde{\alpha}$.
The equations of motion corresponding to $\lambda$
and $\widetilde{\lambda}$ force the field strength $F_{\mu\nu}$ and $\widetilde{F}_{\mu\nu}$
to vanish on the boundary $\dd\cal{M}$. However, these constraints are 
already imposed by the Lagrange multipliers $B_\mu$ and $\widetilde{B}_\mu$.
In this sense, $\lambda$ and $\widetilde{\lambda}$ are redundant
fields and can be set to zero at anytime. In other words, a gauge fixing
of the transformations (\ref{extratrans}) for which $\lambda=\widetilde{\lambda}=0$
can be made. We will nevertheless keep this additional term in our analyses for later use.
Of course we could have also chosen to impose appropriate boundary conditions
on the fields themselves. However, this is not in the spirit of our procedure
as mentioned above.
\par
The Lagrange multiplier terms in the first order action 
are seperately invariant under the gauge transformations (\ref{gaugetrans}).
However, the Chern-Simons parts together with the boundary action 
are not. It is though possible to recover this invariance 
by putting restrictions
on the gauge parameters as given by
\be
L|_{\dd\cal{M}}=R|_{\dd\cal{M}}=0\,\,\,\,
\ee
and demande, for quantum invariance, that the topological charges
$\Gamma\left(L\right)$ and $\Gamma\left(R\right)$ are integer-valued.
\par
As a matter of fact, not all of the gauge symmetry (\ref{gaugetrans})
is exhausted by the gauge fixing conditon $g=1$.  Indeed,  when both $A_\mu$ and $\widetilde{A}$
are present, the first order action is still invariant everywhere under the 
gauge transformations
\be
A_\mu \longrightarrow  HA_\mu H^{-1} - \dd_\mu HH^{-1}
\,\,\,\,\,\,,\,\,\,\,\,\,
\widetilde{A}_\mu \longrightarrow  H
\widetilde{A}_\mu H^{-1} - \dd_\mu HH^{-1}\,\,\,
\ee
with no restrictions on the group element $H$ at the boundary. Accordingly, the Lagrange 
multipliers transform as 
$B_\mu\longrightarrow HB_\mu H^{-1}$ and
$\widetilde{B}_\mu\longrightarrow H\widetilde{B}_\mu H^{-1}$.
Since the Lagrange multiplier terms are each gauge invariant
under this last symmetry, we have therefore a way of making 
two Chern-Simons actions (more precisely their difference) gauge invariant. Namely, by coupling them 
at the boundary in the unique manner as in the first order action
(\ref{first}). This presents an alternative to the usual procedure employed, namely by introducing 
new dynamical fields, in preserving gauge invariance in Chern-Simons theory.
\par
We return now to our main stream to explore duality.  
The idea of duality is not to use the equations of motion of the 
Lagrange multipliers (as these would always lead to the
original WZWN theory) but use instead those of the gauge fields.
In this context, we distinguish two different situations: If the
gauge fields belong to the diagonal or axial subgroups then they appear
at most in a quadratic form in the action $S_{\rm{first}}$.
Hence, they can be completely eliminated from the action through their 
equations of motion (which is equivalent to performing the Gaussian
integration over these fields in the path integral). The resulting
theory (the dual theory) is  another non-linear sigma model 
but with a different target space metric and a different torsion 
from those encountred in the original WZWN model. This case has
been the subject of many investigations \cite{wzwdual}
and will not concern us here.
The second situation, which will be our main interest, deals with 
the gauging of subgroups for which the presence of the Chern-Simons action is required.  
The gauge fields are no longer quadratic in 
the gauged action. Therefore, they cannot
be integrated out from the first order action and the dual
theory is certainly not a non-linear sigma model. 
Let us nevertheless stay at the classical level and 
examin the equations of motion of the gauge fields $A_\mu$ and 
$\widetilde{A}_\mu$. 
\par
The Lie algebras $\cal{G}$ has generators $T_a$ and is specified 
by the structure constants $f^a_{bc}$ and the trace 
$\eta_{ab}={\rm{tr}}\left(T_aT_b\right)$,
where $\eta_{ab}f^b_{cd}+\eta_{cb}f^b_{ad}=0$. 
We do not assume anything 
on the ivertibility of the invariant bilinear form $\eta_{ab}$. The fields of the 
first order action are decomposed as $A_\mu=A^a_\mu T_a$, 
$\widetilde{A}_\mu=\widetilde{A}^a_\mu T_a$,  $B_\mu=B^a_\mu T_a$
and $\widetilde{B}_\mu=\widetilde{B}^a_\mu T_a$. 
We start by calculating the equations of motion in the bulk.
The variation
of the action (\ref{first}) with respect to the gauge fields yields
\be
\epsilon^{\mu\nu\rho}\,\eta_{ab}\left(F^b_{\nu\rho}
 +2{\cal{D}}_\nu B^b_\rho\right)=0
\,\,\,\,\,\,\,\,,\,\,\,\,\,\,\,\,\, 
\epsilon^{\mu\nu\rho}\,\eta_{ab}\left(\widetilde{F}^b_{\nu\rho}
-2\widetilde{{\cal{D}}}_\nu \widetilde{B}^b_\rho\right)=0
\,\,\,\,.
\label{eqofmotion}
\ee
Acting with the covariant derivatives ${\cal D}_\mu$ and $\widetilde{\cal D}_\mu$ 
on these equations and using the 
Bianchi identities, leads to the following consistency relations
\be
\epsilon^{\mu\nu\rho}\,\eta_{ab}\, f^b_{cd} F^c_{\mu\nu}B^d_\rho=0\,\,\,\,\,,\,\,\,\,
\epsilon^{\mu\nu\rho}\,\eta_{ab}\, f^b_{cd} \widetilde{F}^c_{\mu\nu}
\widetilde{B}^d_\rho=0\,\,\,.
\ee
Among the possible solutions to these consistency equations, two are of particular interest to us. 
The first is provided by taking $F^a_{\mu\nu}=\widetilde{F}^a_{\mu\nu}=0$ (we assume 
for this discussion that $\eta_{ab}$ is invertible).
These conditions are as if one is using the equations of motion of the Lagrange
multiplier fields $B_\mu$ and $\widetilde{B}_\mu$. 
Solving these constraints by taking $A_\mu$ and $\widetilde{A}_\mu$ to be pure gauge
fields would eventually yield the original WZWN theory.
Furthermore, the vanishing of the gauge curvatures (and hence the 
disappearance of the Lagrange multiplier terms from the action) are the usual equations
of motion of pure Chern-Simons theory. This explains why
pure Chern-Simons theory (without the BF terms) could be  equivalent, 
in some particular situations, to the WZWN model. 
\par   
The second solution is given by
\be
B^a_\mu=G_{\mu\nu}\,\epsilon^{\nu\alpha\beta}\,F_{\alpha\beta}^a
\,\,\,\,\,\,\,,\,\,\,\,\,\,
\widetilde{B}^a_\mu=G_{\mu\nu}\,\epsilon^{\nu\alpha\beta}\,\widetilde{F}_{\alpha\beta}^a
\,\,\,\,\,,
\ee
where $G_{\mu\nu}$ is the metric on the three dimensional manifold ${\cal M}$.
If we substitute these expressions for $B_\mu$ and $\widetilde{B}_\mu$
in the equations of motion (\ref{eqofmotion}), we obtain
\bea
&&\eta_{ab}\left(\epsilon^{\mu\nu\rho}\,F^b_{\nu\rho}
+2\sqrt{|G|}\, G^{\mu\nu}G^{\alpha\beta}\,{\cal{D}}_\alpha F^b_{\nu\beta}\right)=0
\nonumber\\
&&\eta_{ab}\left(\epsilon^{\mu\nu\rho}\,\widetilde{F}^b_{\nu\rho}
-2\sqrt{|G|}\, G^{\mu\nu}G^{\alpha\beta}\,
\widetilde{{\cal{D}}}_\alpha \widetilde{F}^b_{\nu\beta}\right)=0
\,\,\,\,.
\eea
These are the equations of motion in the bulk of a  
three dimensional Yang-Mills theory in the presence
of Chern-Simons terms. 
This observation might explain the origin of the boundary Kac-Moody algebra
found in a Yang-Mills-Chern-Simons gauge theory \cite{park}. 
Therefore, different dual theories to the WZWN model are obtained
depending on which solution one chooses for the consistency equations.
\par
In order for the above variational procedure to be well-defined, one 
needs to specify the boundary conditions of the problem. 
We choose not to impose by hand boundary conditions on the fields. We rather 
let the equations of motion play their full r\^ole and determine for us 
the boundary conditions. This is achieved by treating the boundary 
variations in the same manner as those of the bulk. In other words, we
demand that $\delta A_\mu$ and $\delta\widetilde{A}_\mu$ are arbitrary both
on the bulk and on the boundary.   
When varying
with respect to $A^a_\mu$ and $\widetilde{A}^a_\mu$, one obtains the boundary
terms
\bea
&&-{k\over{4\pi}}\int_{\dd\cal{M}}{\mathrm{d}}^2x \,
\eta_{ab}\left[P_-^{\rho\mu}\left(\widetilde{A}^a_\mu-A^a_\mu\right)
+2\epsilon^{\rho\mu}\left(B^a_\mu+{\cal D}_\mu\lambda^a\right)\right]
\delta A^b_\rho\nonumber\\
&&-{k\over{4\pi}}\int_{\dd\cal{M}}{\mathrm{d}}^2x \,
\eta_{ab}\left[P_+^{\rho\mu}\left(A^a_\mu-\widetilde{A}^a_\mu\right)
+2\epsilon^{\rho\mu}\left(\widetilde{B}^a_\mu+\widetilde{{\cal D}}\widetilde{\lambda}^a\right)\right]
\delta\widetilde{A}^b_\rho\,\,\,\,,
\eea
where we have included the contribution due to the action (\ref{addaction}) with
$\lambda=\lambda^a T_a$ and $\widetilde{\lambda}=\widetilde{\lambda}^a T_a$.
If we choose 
not to impose the vanishing of $\delta A^a_\mu$ and $\delta \widetilde{A}^a_\mu$
at the boundary, then the equations of motion on the boundary are
\bea
&&
\eta_{ab}\left[P_-^{\rho\mu}\left(\widetilde{A}^a_\mu-A^a_\mu\right)
+2\epsilon^{\rho\mu}\left(B^a_\mu+{\cal D}_\mu\lambda^a\right)\right]
=0\nonumber\\
&&
\eta_{ab}\left[P_+^{\rho\mu}\left(A^a_\mu-\widetilde{A}^a_\mu\right)
+2\epsilon^{\rho\mu}\left(\widetilde{B}^a_\mu +\widetilde{{\cal D}}_\mu
\widetilde{\lambda^a}\right)\right]=0
\,\,\,\,.
\eea
This determines for us, in a natural way, the behaviour of the gauge fields
at the boundary. 
Of course, these boundary conditions must be compatible with 
the bulk equations of motion (\ref{eqofmotion}). Since, the boundary action 
was found in a unique fashion, namely through demanding gauge invariance of the 
WZWN action, we suspect that the two sets of equations are always
compatible. This has been checked at least for the case of the BTZ black hole 
in the absence of BF terms \cite{gaida}.  
\par

\section{Comparaison with three dimensional gravity}

We consider now some special Lie algebras which are relevant to the 
study of three dimensional gravity.
The first of these cases consists in taking the left gauging,  
$g\longrightarrow Lg$,  of the WZWN model. This amounts to setting $\widetilde{A}=0$
in the first order action (\ref{first}). 
We choose a Lie algebra whose generators $T_i$ and $J_i$ satisfy 
\bea
&&\left[T_i\,,\,T_j\right]=f^k_{ij}T_k\,\,\,\,,\,\,\,\,\,
\left[T_i\,,\,J_j\right]=f^k_{ij}J_k\,\,\,\,,\,\,\,\,\,
\left[J_i\,,\,J_j\right]=0
\nonumber\\
&&{\rm tr}\left(T_iT_j\right)=0\,\,\,\,,\,\,\,\,
{\rm tr}\left(J_iJ_j\right)=0\,\,\,\,,\,\,\,\,
{\rm tr}\left(T_iJ_j\right)=\eta_{ij}\,\,\,\,.
\eea
We will work with the modified first order action in (\ref{modfirst})
together with the additional boundary action (\ref{addaction})
and expand the different fields there according to 
\be
A_\mu=\omega_\mu^iT_i+ e_\mu^iJ_i \,\,\,\,,\,\,\,\,\,
Q_\mu=\theta_\mu^iT_i+ v_\mu^iJ_i \,\,\,\,,\,\,\,\,\,
\lambda_\mu=\chi^iT_i+ t^iJ_i \,\,\,\,.
\ee 
The action for this particular Lie algebra takes then the form
\bea
S^{(1)}_{\rm {first}}&=& 
-{k\over{4\pi}}\int_{\cal{M}}{\mathrm{d}}^3y\,\epsilon^{\mu\nu\rho}\,
\eta_{ij}\left[v^i_\mu F^j_{\nu\rho}
+2\theta^i_\mu{\cal D}_\nu e_\rho^j -{1\over 2} f^i_{kl}\omega_\mu^k
\omega^l_\nu e^j_\rho\right]
\nonumber\\
&+& {k\over{4\pi}}\int_{\dd\cal{M}}{\mathrm{d}}^2x\,\eta_{ij}
\left[\sqrt{|\gamma|}\gamma^{\mu\nu}\omega^i_\mu e^j_\nu
-\epsilon^{\mu\nu}\left(t^i F^j_{\mu\nu}
+2\chi^i{\cal D}_\mu e_\nu^j\right)\right]
\,\,\,\,,
\label{example1}
\eea
where $F^i_{\mu\nu}=\dd_\mu\omega_\nu^i -\dd_\nu\omega_\mu^i
+f^i_{jk}\omega^j_\mu\omega^k_\nu$ and 
${\cal D}_\mu e^i_\nu = \dd_\mu e^i_\nu +f^i_{jk}\omega^j_\mu e^k_\nu$. 
We remark that the gauge field component $e^i_\mu$ appears linearly in this action.
It has, therefore, the function of a Lagrange multiplier which imposes the bulk
contraint $\epsilon^{\mu\nu\rho}\,\eta_{ij}\left({\cal{D}}_\nu \theta^j_\rho
-{1\over 4}f^j_{kl}\omega^k_\nu\omega^l_\rho\right)=0$. If a formal solution 
of the form $\omega^i_\mu=O^i_\mu\left(\theta\right)$ exists then the action
(\ref{example1}) is effectively a BF theory with just the first term
present in the bulk. The true fields are, as expected, the original Lagrange multipliers
$\theta^i_\mu$ and $v^i_\mu$. This statement will be of use to us when dealing
with three dimensional gravity.
\par
The other example we consider is obtained for left and right gauging,   
$g\longrightarrow LgR$, of the WZWN model. Here both gauge fields $A_\mu$ and 
$\widetilde{A}_\mu$ are kept in the first order action (\ref{first}).
We assume that the Lie algebra ${\cal G}$ consists of two 
identical copies (left and right) spanned by the generators $T_i$ and
$\widetilde{T}_i$. We have chosen to label the two copies with the 
same indices. The structure constants are denoted $f^i_{jk}$ 
for both copies and the trace is such that ${\rm{tr}}\left(T_iT_j\right)
={\rm{tr}}\left(T_i\widetilde{T}_j\right)=
{\rm{tr}}\left(\widetilde{T}_i\widetilde{T}_j\right)=\eta_{ij}$.
The two independent gauge fields are decomposed according to 
\be
A_\mu=\left(\omega^i_\mu + \alpha \,e_\mu^i\right)T_i\,\,\,\,,\,\,\,\,\,
\widetilde{A}_\mu= \left(\omega^i_\mu - \alpha \,e_\mu^i\right)\widetilde{T}_i
\,\,\,\,.
\ee
Similarly, the redefined Lagrange multipliers in (\ref{modfirst}) are written in the form
\be
Q_\mu=\left(\theta_\mu^i+{1\over \alpha} \,v_\mu^i\right)T_i\,\,\,\,\,\,,\,\,\,\,\,
\widetilde{Q}_\mu=\left(\theta_\mu^i-{1\over \alpha} \,v_\mu^i\right)\widetilde{T}_i\,\,\,\,.
\ee
The action $S_{{\rm first}}$ for this kind of Lie algebra is given by
\bea
S_{\rm{first}}^{(2)}&=&
-{k\over {2\pi}}\int_{\cal{M}}{\mathrm{d}}^3y\,\epsilon^{\mu\nu\rho}\,
\eta_{ij}\left[\theta^i_\mu \left(F^j_{\nu\rho} +\alpha^2
f^j_{kl}e^k_\nu e^l_\rho\right) + 2 v^i_\mu\,{\cal{D}}_\nu e^j_\rho 
\right.\nonumber\\
&-&\left.{1\over 2}\,\alpha\, f^i_{kl}\left(
\omega_\mu^k\omega_\nu^l e^j_\rho + {1\over 3}\,     
\alpha^2 e^k_\mu e^l_\nu e^j_\rho\right)
\right]
+ {k\over{2\pi}}\int_{\dd\cal{M}}{\mathrm{d}}^2x\,\eta_{ij}\left\{\alpha^2\, \sqrt{|\gamma|}
\gamma^{\mu\nu}\,\eta_{ij}\,e 
^i_\mu e^j_\nu\right.
\nonumber\\
&-& \left.
\epsilon^{\mu\nu}\left[\alpha\,\omega^i_\mu e^j_\nu +
\chi^i\left(F^j_{\mu\nu} +\alpha^2\,
f^j_{kl}e^k_\mu e^l_\nu\right) + 2 t^i\,{\cal{D}}_\mu e^j_\nu\right]\right\} 
\,\,\,,
\label{example2}
\eea
where we have also expanded the additional fields in (\ref{addaction}) 
as $\lambda=\left(\chi^i+{1\over\alpha} \,t^i\right)T_i$ and 
$\widetilde{\lambda}=\left(\chi^i-{1\over\alpha} \,t^i\right)\widetilde{T}_i$.
Since the above action 
is cubic in $e^i_\mu$, it is not merely an effective BF theory.
\par
Another reason behind the choice of these particular Lie algebras resides in their
close connection with three dimensional gravity. Before entering into the details
of the precise relashionship, 
let us briefly recall the main features of three dimensional gravity in the Palatini 
formalism (see \cite{carlip2} for a review).
The Einstein-Hilbert action in three dimensions, 
\be
S^{{\rm {EH}}}={1\over 16\pi\kappa}\int_{\cal{M}} {\rm{d}}^3y\,\sqrt{|G|}\,\left(R-2\Lambda\right)
+ \int_{{\dd\cal{M}}} {\rm{d}}^2x\,{\cal L}\left(G\right)\,\,\,\,,
\ee
can be cast in a first order formalism as
\be
S_{\rm{Palatini}}^{\rm{EH}}={1\over 16\pi\kappa}
\int_{\cal{M}}\,\epsilon_{IJK}\,
\left[F^{IJ}\left(\Omega\right)\wedge E^K - {\Lambda\over 3} \,E^I\wedge E^J\wedge E^K
\right] +\int_{{\dd\cal{M}}} {\cal L}\left(\Omega,E\right) \,\,\,\,\,. 
\label{palatini}
\ee
The fundamental variables are a one-form connection $\Omega$ and a one-form triad field $E$.
The exact nature of the boundary term is not relevant to us here and we refer
the reader to \cite{hawking} for more details. The indices
$I,J,K,\dots=0,1,2$ label an internal space whose flat metric we denote $h_{IJ}$
and $\epsilon_{012}=1$.
The spacetime metric is as usual given by $G_{\mu\nu}=h_{IJ}E^I_\mu E^J_\nu$. We take 
$h_{IJ}$ to be  of Lorentzian signature. The curvature two-form is $F^{IJ}\left(\Omega\right)
={\rm d}\Omega^{IJ}+\Omega^I_{\,\,\,K} \wedge\Omega^{KJ}$. It is convenient to identify 
the internal space indices with those of a three dimensional Lie algebra. This is 
$SO(2,1)$ for Lorentzian gravity and $SO(3)$ for a Euclidean spacetime. The internal metric
$h_{IJ}$ is then taken to be proportional to the Killing-Cartan metric of 
this Lie algebra while $\epsilon^I_{\,\,\,JK}=h^{IL}\epsilon_{LJK}$ are its structure constants.
Furthermore, for the connection to take value in the Lie algebra, we 
introduce the new connection (labelled with one index) through the redefinition
$\Omega_I^{\,\,\,J}=\epsilon^J_{\,\,\,IK}\,\Omega^K$. 
The one-form connection
is such that $\Omega^{IJ}=-\Omega^{JI}$ and is, therefore, metric-preserving (it satisfies the
metricity condition).
\par
We return now to our two examples in (\ref{example1}) and (\ref{example2}) and try 
to find their connection to three dimensional gravity. We specialise to the case when 
$\eta_{ij}$ and $f^i_{jk}$ describe an $SO(2,1)$ Lie algebra and are 
proportional to $h_{IJ}$ and $\epsilon^I_{\,\,\,JK}$ respectively. 
We start by examining the first example in (\ref{example1}).
As noticed above, the action 
$S^{(1)}_{{\rm {first}}}$ is effectively a BF theory 
in the bulk and can therefore be identified with $S^{{\rm {EH}}}_{{\rm {Palatini}}}$
with zero cosmological constant. 
The boundary Lagrangian ${\cal{L}}\left(\Omega,E\right)$ is unique and corresponds to that
of $S^{(1)}_{{\rm {first}}}$.
The r\^ole of the triads $E^I_\mu$ is played by the field
$v^i_\mu$ and 
the three dimensional metric is then given by $G_{\mu\nu}=\eta_{ij}v^i_\mu v^j_\nu$.
On the other hand, the connection $\Omega^I_\mu$ is related to $O^i_\mu\left(\theta\right)$;
the solution to the constraints imposed by $e^i_\mu$ on the fields of the action
(\ref{example1}). 
There is, however, a crucial difference between the two actions
if one interprets $\omega^i_\mu$ (and not $O^i_\mu\left(\theta\right)$)
as the spin connection of three dimensional gravity: The variation of (\ref{palatini})
with respect 
to $\Omega^I_\mu$ implies a torsion-free condition whereas a variation of (\ref{example1})
with respect to
$\omega^i_\mu$ does not. This is due to the 
presence of the last two terms in the bulk part of the action (\ref{example1}).
In addition, there are more fields in (\ref{example1}) than in (\ref{palatini}) 
and the two theories coincide only if all the Lagrange multipliers in (\ref{first}) are set to zero.
This amounts to setting $\theta^i_\mu={1\over 2}\omega^i_\mu$ and $v^i_\mu={1\over 2}e^i_\mu$
in the action (\ref{example1}).
\par
The situation with the action $S^{{{(2)}}}_{{\rm {first}}}$ is much more complex.
If the Lagrange multipliers were absent then our action in
(\ref{example2}) reduces to $S^{{\rm {EH}}}_{{\rm {palatini}}}$ in the presence
of a non vanishing cosmological constant. This can be seen by setting 
$\theta^i_\mu={1\over 2}\,\alpha e^i_\mu$ and $v^i_\mu={1\over 2}\,\alpha\omega^i_\mu$
together with performing an integration by parts in the action (\ref{example2}).  
In general, however, the two actions are different. 
This is mainly due to the fact that the field $e^i_\mu$ is cubic in the action 
(\ref{example2}) and cannot be integrated out. This makes it difficult to determine the
fields that play the r\^ole of the triads and the connection of three dimensional
gravity with a non vanishing cosmological constant.

\section{Conclusions}

The WZWN model is a conformal field theory that has been used in the past to 
reproduce various two dimensional theories like Toda field theories,
black holes and others. In this paper we have enlarged this liste by connecting
this model to a combination of three dimensional topological BF and Chern-Simons gauge theories defined
on a manifold with boundaries.  In order for this connection to hold, the boundary action 
accompanying these topological theories is unique. We arrive to this result by a direct
application of non Abelian duality on the WZWN non-linear sigma model.  
One of our motivations in this work is to provide, at the classical level and at the level 
of the Lagrangian, a precise relashionship between the WZWN model and three dimensional
gravity. This is a special case in our study an two examples are supplied. 
In the case of the Lie algebra $SO\left(2,1\right)$, we find that one obtains 
three dimensional gravity without a cosmological constant. However,  
the triads and the connection of gravity are not simply given by the components of the 
Chern-Simons gauge field.  The other example is based on the Lie algebra 
$SO\left(2,1\right)\times SO\left(2,1\right)$ and yields  three dimensional 
gravity with a cosmological constant only if the BF theory is not taken
into account.
\par
As we have stressed before, the integration over the Lagrange multipliers at any time
would yield the original WZWN model. 
Notice that in both of our examples, the equations of motion of the Lagrange 
multipliers are simply Einstein's equations (with and without cosmological constant) and the torsion-free
condition, if one interprest $e^i_\mu$ and $\omega^i_\mu$, respectively, as the triads and 
connection of tree dimensional gravity. Therefore, it is 
{\it {only on-shell}} (that is, when Einstein's equations and the torsion-free condition are 
satisfied) that three-dimensional gravity, with our unique boundary action, is equivalent to the
WZWN theory. This statement is, of course, equivalent to ignoring the BF contribution.   
\par
One of the interesting problems would be of course to quantise the resulting BF and Cherns-Simons theories
taking into account the boundary terms. This might provide a map between the observables of the two
dual theories, namely the WZWN model and the topological theory. In the absence of boundaries, a
quantum treatment of a combined BF and Chern-Simons has been presented in \cite{yahikosawa}.
There also, the question of which fields might play the r\^ole of the  triads and connection of 
gravity is raised. We should also mention that we are still at the level the first order action. 
The next step in the completion of the dualisation procedure is to perform the integration over
the gauge field. These appear in a cubic form in the Lagrangian and no known methods allow
their complete integration from the path integral. However, a perturbative analyses is always
possible. Notice though that there are situations, like the example in (\ref{example1}), 
where the gauge components appear at most in a quadratic form and without any derivatives 
acting on them. Integrating them out from the path integral is, in principle, feasible.
It might be instructive to start by the quantisation of such models.

\par
\noindent
{\bf {Aknowledgements}}: 
I would like to thank J. Balog and P. Forg\'acs for discussions.

\end{document}